\documentclass[aps,pra,floatfix,superscriptaddress,twocolumn,showpacs,10pt]{revtex4-1}
\usepackage{amssymb, amsmath,color,mciteplus,graphicx,subfigure}
\usepackage{calc,epsfig,epstopdf,color,mciteplus,bm,mathrsfs}
\usepackage{times}
\usepackage{float}
\usepackage{multirow}
\usepackage{lipsum}

\begin{document}

\title{State-independent geometric quantum gates via nonadiabatic and noncyclic evolution}

\author{Yue Chen}

\author{Li-Na Ji}
\affiliation{Key Laboratory of Atomic and Subatomic Structure and Quantum Control (Ministry of Education), and School of Physics, South China Normal University, Guangzhou 510006, China}

\author{Zheng-Yuan Xue}\email{zyxue83@163.com}
\affiliation{Key Laboratory of Atomic and Subatomic Structure and Quantum Control (Ministry of Education), and School of Physics, South China Normal University, Guangzhou 510006, China}
\affiliation{Guangdong Provincial Key Laboratory of Quantum Engineering and Quantum Materials, Guangdong-Hong Kong Joint Laboratory of Quantum Matter,  and Frontier Research Institute for Physics,\\ South China Normal University, Guangzhou 510006, China}

\author{Yan Liang}\email{liangyan9009@163.com}
\affiliation{Key Laboratory of Atomic and Subatomic Structure and Quantum Control (Ministry of Education), and School of Physics, South China Normal University, Guangzhou 510006, China}
\affiliation{College of Physics Science and Technology, Guangxi Normal University, Guilin 541004, China}
\date{\today}

\begin{abstract}
Geometric phases are robust to local noises and the nonadiabatic ones can reduce the evolution time, thus nonadiabatic geometric gates have strong robustness and can approach high fidelity. However, the advantage of geometric phase has not being fully explored in previous investigations. Here, we propose a scheme for universal quantum gates with pure  nonadiabatic and  noncyclic geometric phases from  smooth evolution paths. In our scheme, only   geometric phase can be accumulated in a fast way, and thus it not only fully utilizes the local noise resistant property of geometric phase but also reduces the  difficulty in experimental realization. Numerical results show that the implemented geometric gates  have stronger robustness than dynamical gates and the geometric scheme with cyclic path. Furthermore, we propose to  construct universal quantum  gate on  superconducting circuits, with the fidelities of single-qubit gate and nontrivial two-qubit gate can achieve $99.97\%$ and $99.87\%$, respectively. Therefore, these high-fidelity quantum gates are promising  for large-scale fault-tolerant quantum computation.

\end{abstract}

\maketitle

\section{Introduction}

Compared with classical computation, quantum computation has the intrinsic potential to deal with tasks in a  parallel way,  due to its unique quantum superposition  property \cite{Nielson}, which can exponentially speed-up quantum computers. Therefore, quantum computation can effectively deal with hard problems \cite{Grover,Shor} that classical computation cannot solve.  
 It is well-known that building high-fidelity  quantum control is indispensable in various fields of quantum technology  \cite{ Lloyd, Yliang, YanB, ZhangJ}. 
However, it is impossible  to avoid the decoherence effects from the surrounding environment throughout the entire computation process, leading to a degradation of the gate fidelity. In order to achieve large-scale and fault-tolerant quantum computation, we need to find methods to reduce local operational errors and the influence of decoherence to obtain   high-fidelity and robust quantum gates.

Berry \cite{Berry} noticed  that in quantum systems, quantum states will also acquire a geometric phase, together with the dynamical phase, after an adiabatic and cyclic evolution. Its unique global properties, unaffected by the details of evolution, enhances its ability  in resisting  local parameter noise during evolution, leading to the development of geometric quantum computation. %Later, its non-Abelian generalization \cite{Wilczek} was developed. 
However, %in both cases, 
the needed evolution time is too long due to the constrain of  the adiabatic condition, which considerably  impact the gate fidelity greatly. Subsequently, Aharonov and Anandan \cite{Aharonov} found  that the adiabatic condition is not necessary, as long as certain requirements are met. This paved the way for %Abelian  and non-Abelian \cite{Z. Y. Xue,P. Z. Zhao,G. F. Xu,E. Sjoqvist} 
geometric quantum computation based on the nonadiabatic evolution, i.e., nonadiabatic geometric quantum computation (NGQC)   \cite{wangxb, zhuwang, Zhu, Chen2020, Ding, XuG}, %with Abelian geometric phases has been successfully 
with experimental demonstrations  in many quantum systems, such as trapped ions \cite{Wineland, Ai, Aimz}, NV center in diamond \cite{Zu,Sekiguchi}, nuclear magnetic resonance \cite{Du,Feng, LiY}  and superconducting quantum circuits \cite{Xu,Abdumalikov}, etc.

The geometric gate is more robust than dynamical gate (DG) because it can resist local noise well \cite{ Solinas,Lupo}. In the existing NGQC schemes, the famous orange-slice-shaped-loop scheme \cite{Ota, Thomas,  Zhao, Chen2018} uses a geodesic path to make the dynamical phase zero in the whole process to construct the geometric gate. However, the scheme cannot make arbitrary initial states remain only geometric phase, which affects the fidelity of the gate. The appearance of state-independent nonadiabatic geometric quantum computation (SINGQC) scheme \cite{Liang} resolves this problem, which ensures that the dynamical phase is zero for any initial state, greatly enhancing the robustness of geometric gate. However, the SINGQC scheme adopts cyclic evolution, and the evolution path is divided into several parts, which  increases the difficulty of experimental realization and  the lengthy evolution time there also affect the gate fidelity.

To avoid these drawbacks, we propose the nonadiabatic pure geometric quantum computation (NPGQC) scheme, which focuses on ensuring that any initial states accumulate zero dynamical phase to achieve nonadiabatic pure geometric gates. Besides, we select a noncyclic smooth path, which can shorten the evolution path to reduce evolution time and  also decrease experimental complexity. Through numerical simulations, we find that compared with DG and SINGQC schemes, geometric gates in our scheme are more robust. In addition, our scheme can be implemented in various physical systems. Here, we complete the construction of single-qubit and nontrivial two-qubit geometric gates on the superconducting quantum circuits. Numerical results show that the fidelity of single-qubit gate can reach $99.97\%$ and the fidelity of two-qubit gate is as high as $99.87\%$. Therefore, our scheme provides a promising alternative for large-scale fault-tolerant quantum computation.

\section{The protocol}
In this section, we first calculate the condition of realizing the NPGQC through a noncyclic path. Then, we give the specific method for constructing single-qubit geometric quantum gates. Finally, we test the gate fidelity and compare it with the SINGQC  and DG schemes.

\subsection{The NPGQC scheme}
In our scheme, we adopt the reverse engineering of the target Hamiltonian \cite{Chen,Kang} to construct single-qubit geometric quantum gates. Since there are some conditions to be met, we need to increase the degree of freedom of the auxiliary vectors. Here, we define a pair of orthogonal basis with multiple degrees of freedom of the two-level system as $|\mu_{1}(t)\rangle=\cos\frac{\Gamma}{2}e^{-{\rm i}\frac{\xi}{2}}|0\rangle +\sin\frac{\Gamma}{2}e^{{\rm i}\frac{\xi}{2}}|1\rangle$, and $|\mu_{2}(t)\rangle=\sin\frac{\Gamma}{2}e^{-{\rm i}\frac{\xi}{2}}|0\rangle-\cos\frac{\Gamma}{2}e^{{\rm i}\frac{\xi}{2}}|1\rangle$, where $\Gamma$ and $\xi$ are time-independent parameters. The orthogonal auxiliary vectors $|\psi_{1}(t)\rangle$ and $|\psi_{2}(t)\rangle$ can be designed as 
\begin{eqnarray}
\label{3}
|\psi_{1}(t)\rangle=\cos\frac{\theta}{2}e^{-\rm{i}\frac{\varphi}{2}}|\mu_{1}(t)\rangle +\sin\frac{\theta}{2}e^{\rm{i}\frac{\varphi}{2}}|\mu_{2}(t)\rangle, \notag\\
|\psi_{2}(t)\rangle=\sin\frac{\theta}{2}e^{-\rm{i}\frac{\varphi}{2}}|\mu_{1}(t)\rangle-\cos\frac{\theta}{2}e^{\rm{i}\frac{\varphi}{2}}|\mu_{2}(t)\rangle,
\end{eqnarray}
where time-dependent parameters $\theta$ and $\varphi$ are the polar and azimuth angle of Bloch sphere, respectively.

Then we can parameterize the evolution states $|\varPhi_{k}(t)\rangle\left(k=1,2\right)$ as $|\varPhi_{k}(t)\rangle=e^{{\rm i}\gamma_{k}(t)}|\psi_{k}(t)\rangle$, which satisfy the Schr\"{o}dinger equation $\mathcal{H}(t)|\varPhi_{k}(t)\rangle={\rm i}|\dot{\varPhi}_{k}(t)\rangle$. $\gamma_{k}(t)$ is the total phase with $\gamma_{k}(0)=0$. In addition, when $\gamma_{k}(t)={\rm i}\int_{0}^{t}\langle\psi_{k}(t')|\dot{\psi}_{k}(t')\rangle dt'$, $|\varPhi_{k}(t)\rangle$ can obtain a pure geometric phase in the whole process of evolution. Further $\gamma(t)=\gamma_{1}(t)=-\gamma_{2}(t)=\frac{1}{2}\int_{0}^{t}\dot{\varphi}\cos\theta dt'$ can be obtained. In this way, we can get the Hamiltonian  as \cite{ Li}
\begin{equation}
\label{4}
\begin{split}
\mathcal{H}(t)=&{\rm i}\sum^2_{k\neq l}\langle\psi_{l}(t)|\dot{\psi}_{k}(t)\rangle|\psi_{l}(t)\rangle\langle\psi_{k}(t)|  \\
     =&\frac{1}{2}\left(
  \begin{array}{cc}
    -\Delta(t)& \Omega(t)\\
   \Omega^*(t)& \Delta(t)\\
\end{array}\right),\\
\end{split}
\end{equation}
where
\begin{eqnarray}
\label{9}
\begin{split}
\Delta(t)=&\left(\frac{1}{2}\cos\varphi\sin2\theta\sin\Gamma-\cos \Gamma\sin^2\theta\right)\dot{\varphi}\\
+&\dot{\theta}\sin\Gamma\sin\varphi,\\
\Omega(t)=&\{\frac{1}{4}e^{-\rm{\rm{i}}(\xi+\varphi)}\sin2\theta[1+e^{2i\varphi}
(-1+\cos\Gamma)+\cos\Gamma]\\
+&e^{-\rm{\rm{i}}\xi}\sin\Gamma\sin^2\theta\}\dot{\varphi}+ e^{-\rm{i}\xi}(\rm{i}\cos\varphi+\cos\Gamma\sin\varphi)\dot{\theta},
\end{split}
\end{eqnarray}
with $\Delta(t)$ being the detuning, and $\Omega(t)$ being the driving strength. Then, the corresponding evolution operator is  
\begin{equation}
\label{06}
\begin{split}
U(\tau)=&\!\!\!\sum^2_{k=1}e^{{\rm i}\gamma_{k}(\tau)}|\psi_k(\tau)\rangle\langle\psi_{k}(0)|\\
=&\left(
  \begin{array}{cc}
    u_1& u_2\\
   -u_2^*& u_1^*\\
  \end{array}\right),\\
\end{split}
\end{equation}
where
\begin{equation}\label{006}
\begin{split}
 u_1=&\frac{1}{2}e^{-\frac{1}{2}{\rm i}(\varphi_{+}+2\gamma)}\{\sin\frac{\theta_0}{2}[-\cos\frac{\theta_{\tau}}{2}\sin\Gamma(-1+e^{\rm{i}(\varphi_{+}+2\gamma)}\\
 +&2e^{{\rm i}(\varphi_{0}+2\gamma)}\cos^2\frac{\Gamma}{2}\sin\frac{\theta_{\tau}}{2})+2e^{{\rm i}\varphi_{\tau}}\sin\frac{\theta_{\tau}}{2}\sin^2\frac{\Gamma}{2}]\\
 +&\cos\frac{\theta_0}{2}[2e^{\rm{i}(\varphi_{\tau}+2\gamma)}\cos\frac{\theta_{\tau}}{2}\sin^2\frac{\Gamma}{2}+2e^{{\rm i}\varphi_0}\cos\frac{\theta_{\tau}}{2}\cos^2\frac{\Gamma}{2}\\
 +&(e^{{\rm i}\varphi_{+}}-e^{2{\rm i}\gamma})\sin\frac{\theta_{\tau}}{2}\sin\Gamma]\},\\
 u_2=&\frac{1}{2}e^{-\frac{1}{2}{\rm i}[\varphi_{+}+2(\gamma+\xi)]}\{-\cos\frac{\theta_{\tau}}{2}[2e^{{\rm i}(\varphi_{+}+2\gamma)}\sin\frac{\theta_0}{2}\sin^2\frac{\Gamma}{2}\\
 +&2\cos^2\frac{\Gamma}{2}\sin\frac{\theta_0}{2}-(e^{{\rm i}\varphi_0}-e^{\rm{i}(\varphi_{\tau}+2\gamma)})\cos\frac{\theta_0}{2}\sin\Gamma]\\
 +&\sin\frac{\theta_{\tau}}{2}[2e^{{\rm i}\varphi_{+}}\cos\frac{\theta_0}{2}\sin^2\frac{\Gamma}{2}+2e^{2{\rm i}\gamma}\cos\frac{\theta_0}{2}\cos^2\frac{\Gamma}{2}\\
 -&(e^{{\rm i}\varphi_{\tau}}-e^{{\rm i}(\varphi_0+2\gamma)})\sin\frac{\theta_0}{2}\sin\Gamma]\},
\end{split}
\end{equation}
with $\varphi_{+}=\varphi_{\tau}+\varphi_{0}$, where $\varphi_{0}=\varphi(0)$ and $\varphi_{\tau}=\varphi(\tau)$. The evolution operator in Eq. (\ref{06}) is a nonadiabatic noncyclic geometric quantum gate, with $\gamma_k(\tau)$ being the nonadiabatic and noncyclic geometric phase. When the cyclic condition $|\psi_{k}(\tau)\rangle=|\psi_{k}(0)\rangle$ is satisfied, the evolution operator reduces to the traditional nonadiabatic geometric quantum gate, with $\gamma_{k}(\tau)$ becoming the nonadiabatic cyclic geometric phase.

\begin{figure}[tb]
  \centering
  \includegraphics[width= \linewidth]{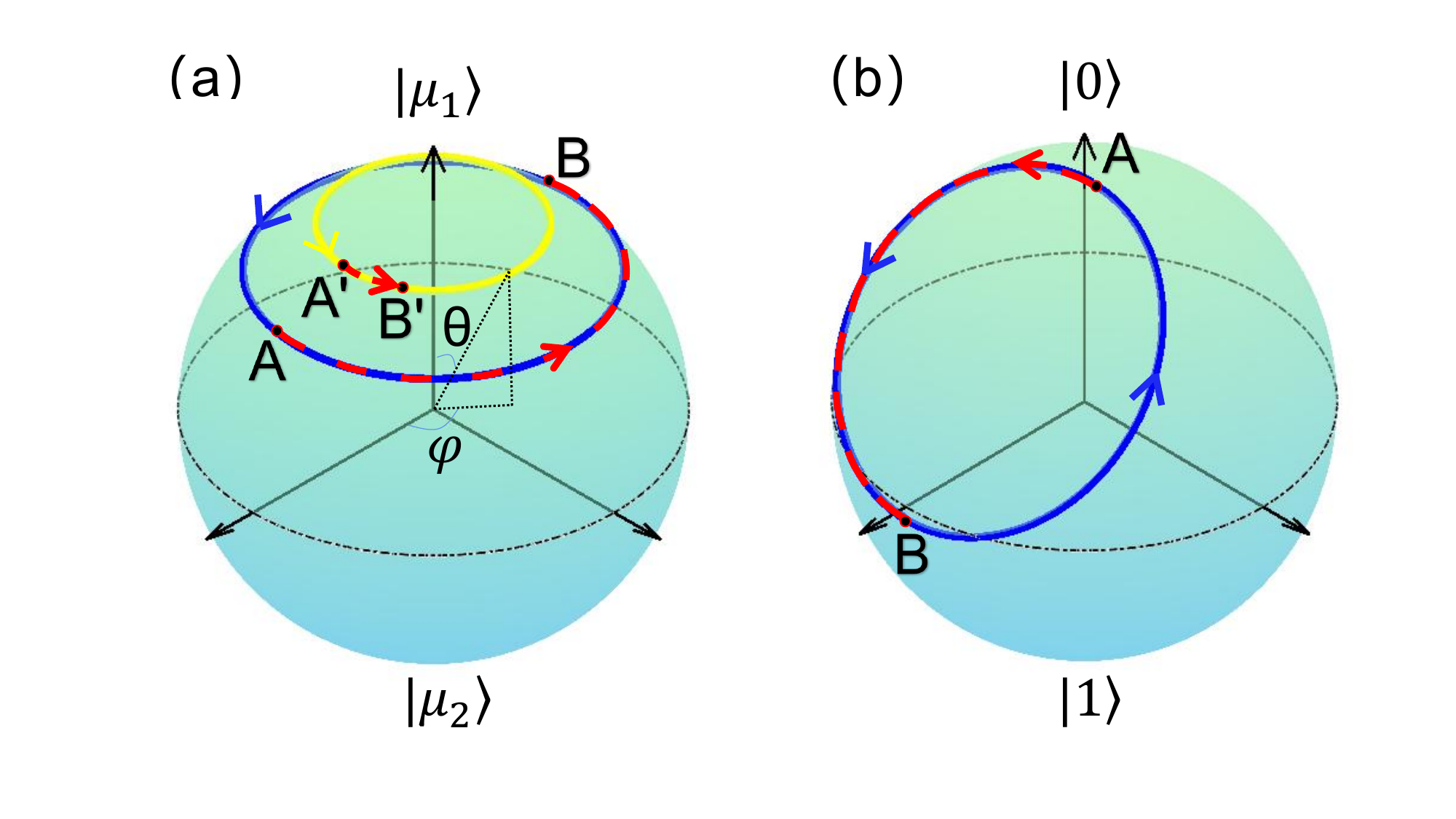}
\caption{ A(A$'$) and B(B$'$) are the starting and ending points of the evolution path. The yellow and blue path denote the trajectories of $T$ and $H$ gates with state $|\psi_1(t)\rangle$, respectively. And the red section is the repeating part. (a) The evolution trajectory in $\{|\mu_1\rangle,|\mu_2\rangle\}$ space, where $\theta$ and $\varphi$ correspond to the polar and azimuth angle of Bloch sphere, respectively. (b) The evolution trajectory in $\{|0\rangle,|1\rangle\}$ space. The trajectory of $T$ gate is the same as the yellow line in (a), thus not presented here.  }\label{fig1} 
\end{figure}

However, for a general evolution state $|\Psi(t)\rangle=e^{-{\rm i}\zeta}\cos\frac{\Lambda}{2}|\varPhi_{1}(t)\rangle+\sin\frac{\Lambda}{2}|\varPhi_{2}(t)\rangle$, where $\Lambda$ and $\zeta$ are arbitrary time-independent parameters, the accumulated dynamic phase during the evolution is no longer zero, but
\begin{eqnarray}
\label{5}
&&\gamma_d(\tau) =\int_{0}^{\tau} \langle\Psi (t)|\mathcal{H}(t)|\Psi (t)\rangle dt \\
    &&= \frac{1}{2}\int_{0}^{\tau}\sin{\Lambda}[\cos(\zeta+2\gamma)\dot{\varphi}\sin\theta-\sin(\zeta+2\gamma)\dot{\theta}] dt.  \notag 
\end{eqnarray}
The condition for an arbitrary evolution state to only accumulate geometric phase at the final time is $\gamma_{d}(\tau)=0$.

\subsection{Single-qubit geometric gates}

There are many evolution pahts that can  satisfy the above condition, here, we adopt one of them to construct the geometric gates, i.e., $\dot{\theta}=0$. In this way, $\gamma(t)=\frac{1}{2}\cos\theta(\varphi-\varphi_0)$, and Eq. (\ref{5}) can be reduced to
\begin{equation}
\label{11}
\gamma_d(\tau)=\frac{1}{2}\sin\Lambda\sin\theta\int_{0}^{\tau}\cos[\zeta+\cos\theta(\varphi-\varphi_0)]\dot{\varphi} dt,
\end{equation}
when $\varphi=\frac{2\pi t}{\tau\cos\theta }+\varphi_{0}$, the dynamical phase can be eliminated. Then  the total phase  is   geometric, and it is
\begin{equation}
\label{12}
\gamma(\tau)=\frac{1}{2}\cos\theta(\varphi_{\tau}-\varphi_0) \\
=\frac{1}{2}\cos\theta\frac{2\pi}{\cos\theta} \\
=\pi.
\end{equation}
The elements of corresponding evolution operator in Eq. (\ref{006}) can also be simplified to
\begin{equation}
\label{12}
\begin{split}
 u_1=& -\cos\frac{\varphi_{-}}{2}+{{\rm i}}\cos\Gamma\sin\frac{\varphi_{-}}{2},\\
 u_2=& \sin\Gamma\sin\frac{\varphi_{-}}{2}\left(\sin\xi+{{\rm i}}\cos\xi\right),
\end{split}
\end{equation}
where $\varphi_{-}=\varphi_{\tau}-\varphi_0$. Therefore, we can construct single-qubit geometric quantum gates via setting the parameter of $(\Gamma,\xi,\varphi_{-})$ and determining the specific form of the Hamiltonian in Eq. (\ref{9}). By setting the parameter of $(\Gamma,\ \xi,\ \varphi_{-})$ as $(\pi/4,\ 0, \ 3\pi)$, $(0,\ 0,\  9\pi/4)$ and $(0,\ 0,\ 5\pi/2)$, the $H$, $T$ and $S$ gates can be obtained, respectively, which can be used to achieve universal single-qubit gates. In order to better demonstrate the advantages of this scheme, $H$ gate and $T$ gate will be selected for testing. Since $S$ and $T$ gates are similar, the $S$ gate does not show here. It is worth noting that with the corresponding expression of $|\psi_{k}(t)\rangle\left(k=1,2\right)$, the evolution path of geometric phase gates can be clearly demonstrated in $\{|0\rangle,|1\rangle\}$ and $\{|\mu_1\rangle,|\mu_2\rangle\}$ space on Bloch sphere, respectively. It can be seen that all of them are smooth routes without segment in the middle, as illustrated in Fig. 1.

\subsection{Gate performance}

Due to the inevitable influence of the surrounding environment, the decoherence effect need to be considered in the performance test of geometric gates. First, to numerically simulate the destruction of decoherence on the robustness of gates, we use the Lindblad master equation \cite{Lindblad} of 
\begin{eqnarray}
\label{EqMaster}
\dot\rho_1&=&-{\rm i}[\mathcal{H}(t), \rho_1]+\kappa_{1}\chi(c_{-})+\kappa_{2}\chi(c_{z}),
\end{eqnarray}
where $\rho_1$ is the density operator, $\kappa_{1}$ and $\kappa_{2}$ respectively denote decay and dephasing rates, the Lindblad operator $\chi(c)=2c\rho c^{+}-c^{+}c\rho-\rho c^{+}c$, with decay operator $c_{-}=|0\rangle\langle1|$ and dephasing operator $c_{z}=|0\rangle\langle0|-|1\rangle\langle1|$. When testing the two geometric gates, $|\Psi(0)\rangle=\cos\Theta|0\rangle+\sin\Theta|1\rangle$ is selected as the initial state, and after the evolution of $H$ and $T$ gates, the final state becomes $|\Psi(\tau)\rangle=[\left(\cos\Theta+\sin\Theta\right)|0\rangle+\left(\cos\Theta-\sin\Theta\right)|1\rangle]/\sqrt{2}$ and $\cos\Theta|0\rangle+\rm{exp}({\rm i}\pi/4)\sin\Theta|1\rangle$, respectively. The gate fidelity is defined as \cite{Poyatos} $F_1=\int_0^{2\pi}\langle \Psi_{\tau}|\rho_1|\Psi_{\tau}\rangle d\Theta/(2\pi)$, which is numerical integrated by the 1001 initial states, with $\Theta$ being uniformed  distribute within $[0, 2\pi]$.

In order to demonstrate that our geometric gate scheme has stronger robustness and higher fidelity than DG scheme, we select a typical dynamical gate as reference for gate performance test. Generally, constructing the Hamiltonian in Eq. (\ref{9}) with $\Delta=0$ and $\Omega(t)=\Omega’(t)e^{-\rm{i}\phi_d}$, we can get  dynamical evolution operator as 
\begin{equation}\label{D1}
\begin{split}
&U_d(\vartheta_d, \phi_d)=e^{-\rm{i}\int_0^\tau\mathcal{H}(t)dt}\\
 &=\left(
  \begin{array}{cc}
  \cos(\frac{\vartheta_d}{2})&-{\rm i}\sin(\frac{\vartheta_d}{2})e^{-{\rm i}\phi_d}\\
   -{\rm i}\sin(\frac{\vartheta_d}{2})e^{{\rm i}\phi_d}&\cos(\frac{\vartheta_d}{2})\\
  \end{array}\right),
\end{split}
\end{equation}
where $\vartheta_d\!=\!\int_0^\tau\Omega'(t)dt=\int_0^\tau\Omega_{\rm m}\sin(\pi t/\tau)dt$. %and constant $\phi_d$ ensures the geometric phase is zero. 
Then, the $H$ gate can be realized by $U_d(\pi, \pi)U_d(\pi/2, \pi/2)$. Besides, $T$ and $S$ gates can be obtained by $U_d(\pi/2, \pi)U_d(\vartheta_z,-\pi/2)U_d(\pi/2, 0)$
 with $\vartheta_z=\pi/2$ and $\pi/4$, respectively.

\begin{figure}[t]
  \centering
  \includegraphics[width=1\linewidth]{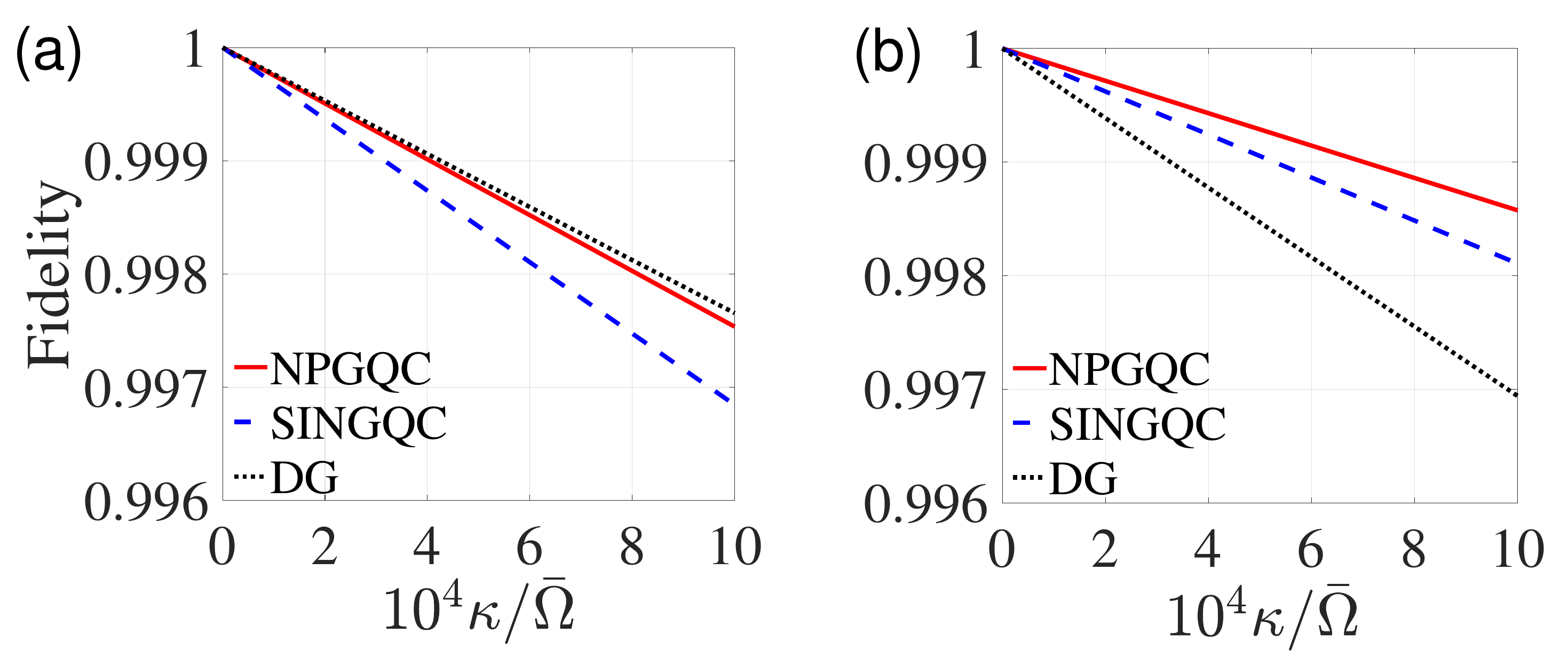}
\caption{  The performance of (a) $H$ gate and (b) $T$ gate in terms of gate-fidelity $F_1$ with respective to decoherence with rate $\kappa$, in unit of $(\bar{\Omega} / 10^4)$, for our scheme, SINGQC scheme and DG scheme.
}\label{fig2}
\end{figure}
 
Then, we plot the results in Fig. 2, where $\bar{\Omega}$ is the average value of driving field amplitude, demonstrating that the fidelity of this scheme, SINGQC scheme and DG scheme vary with different decoherence rates. It is obvious that the fidelity of $T$ gate in this scheme is higher than that of the other two schemes. Furthermore, the gate fidelity of $H$ gate does not exceed the DG scheme, but is much higher than the SINGQC scheme and approaches to the DG scheme. 

In addition to the effect of decoherence on gate fidelity, there are also some systematic errors to be taken into account. Here, we add the qubit frequency drift $\delta$ in the z direction and the driving intensity control error $\epsilon$ in the x direction to the $H$ and $T$ gates, and the Hamiltonian of the system becomes
\begin{equation}
\label{19}
\begin{split}
\mathcal{H}(t)=&\frac{1}{2}\left(
  \begin{array}{cc}
    -[\Delta(t)+\delta\bar{\Omega}]& (1+\epsilon)\Omega(t)\\
   (1+\epsilon)\Omega^*(t)& \Delta(t)+\delta\bar{\Omega}\\
\end{array}\right).\\
\end{split}
\end{equation}
Fig. 3 is a three-dimensional plot of gate fidelity varying with $\delta$ and $\epsilon$ errors for $\kappa_{1}=\kappa_{2}=2 \bar{\Omega}/10^{4}$. It is worth noting that compared with SINGQC and DG schemes, our scheme has a larger range of gate fidelity above $99.9\%$ in the picture under the same error value variation, so the gate of our scheme has the strongest robustness.

\begin{figure}[tbp]
  \centering
\includegraphics[width=\linewidth]{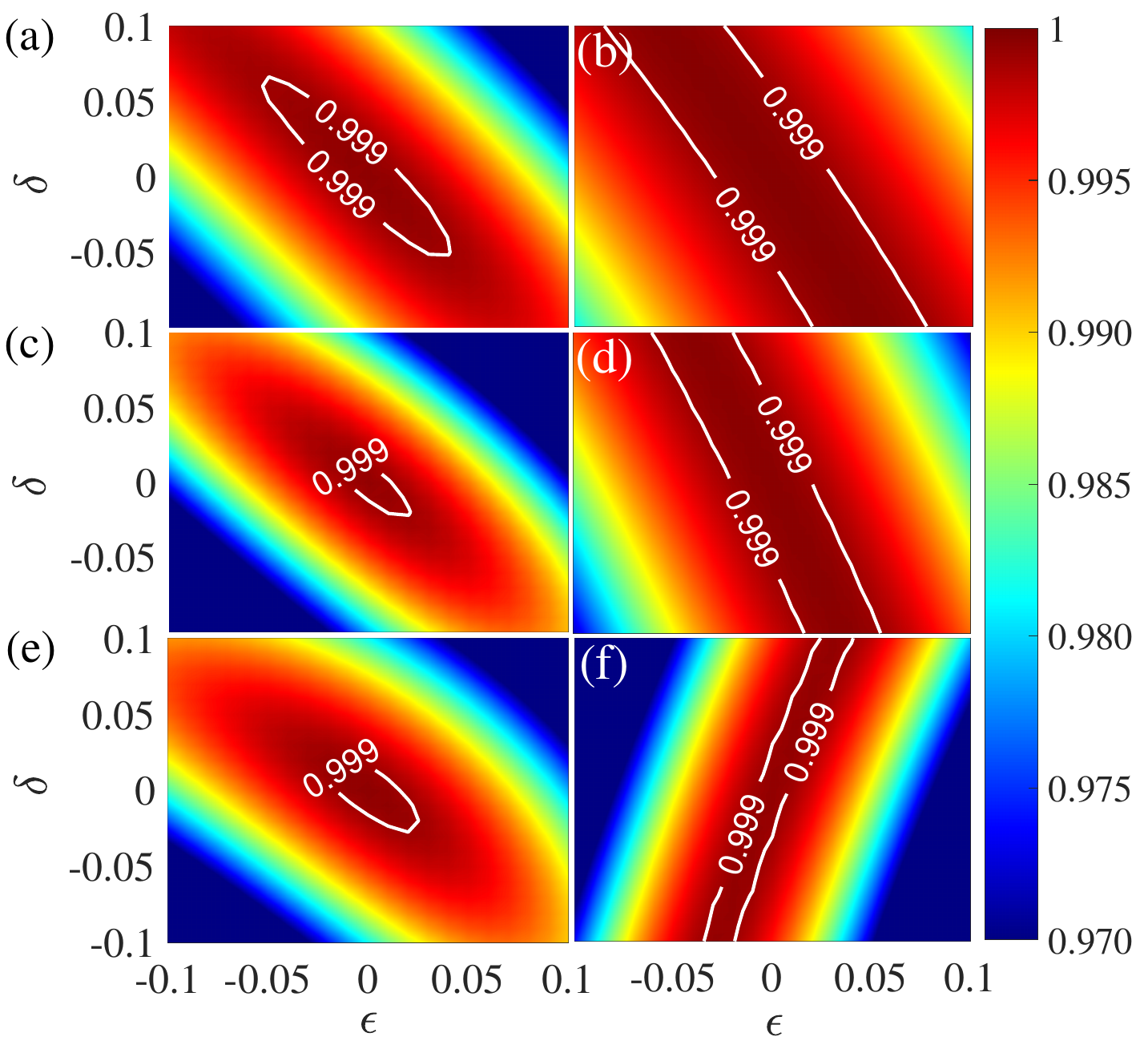}
\caption{ The robustness of $H$ gate in (a) our scheme, (c) SINGQC scheme and (e) DG scheme with respective to pulse-control and qubit-frequency-drift errors, represented by the error fraction of $\epsilon$ and $\delta$, respectively. The  robustness of $T$ gate in (b) our scheme, (d) SINGQC scheme and (f) DG scheme with respective to $\epsilon$ and $\delta$ error fractions, respectively. And the decoherence rates are chosen as $\kappa_{1}=\kappa_{2}=2 \bar{\Omega}/10^{4}$.
}\label{fig3}
\end{figure}

\section{ Physical realization}\label{99}
Here, we construct single-qubit geometric gate and nontrivial two-qubit geometric gates in a capacitance-coupled superconducting transmon qubits \cite{You, transmon}, and Fig. 4(a) is a 2D square lattice in which all adjacent transmon are capacitatively coupled. Furthermore, we test the performance of the gate through numerical simulation by combining the currently available experimental techniques.

\subsection{Universal single-qubit gate}
At first, we implement single-qubit gates in superconducting systems. The energy spectrum for a transmon qubit is shown in Fig. 4(b). Driven by a microwave field, the qubit information will leak  to the second excited state or even higher excited states due to the weak anharmonicity $\alpha$ of the transmon qubit, which leads to the leakage error on the quantum gates. Therefore,  we adopt the "derivative removal via adiabatic gate" (DRAG) technology \cite{Gambetta,Wang} to supress the leakage error. We consider the leakage from the qubit space to the second excited state, which is the main source of the leakage, and the Hamiltonian of this single-qubit system can be written as
\begin{eqnarray}
\label{67}
\mathcal{H}_{D}(t)&=&\frac{1}{2}[\mathbf{B}_{0}(t)+\mathbf{B}_{d}(t)]\cdot \mathbf{S}-\alpha|2\rangle\langle2|,
\end{eqnarray}
where $\mathbf{B}_0(t)$ and $\mathbf{B}_d(t)$ are the vectors of original and DRAG corrected microwave fields, they can be respectively written as
\begin{eqnarray}
\label{52}
\mathbf{B}_{0}(t)&=&
\begin{cases}
B_{x}(t) =\Omega'(t)\cos(\phi_0+\phi(t)), \\
    
    B_y(t) =\Omega'(t)\sin(\phi_0+\phi(t)),  \\     
    B_z(t) =-\Delta(t),\\
\end{cases} \\
\mathbf{B}_{d}(t)&=&
\begin{cases}
B_{dx}(t) =\frac{1}{2\alpha}\left(\dot{B}_y(t)-B_z(t)B_x(t)\right), \\
    
    B_{dy}(t) =-\frac{1}{2\alpha}\left(\dot{B}_x(t)+B_z(t)B_y(t)\right),  \\     
    B_{dz}(t) =0,\\
\end{cases}
\end{eqnarray}
where $\Omega'(t)$ and $\phi(t)$ are driving strength and phase, respectively. Besides, the component vectors of $\mathbf{S}$ in the $x$, $y$, and $z$ directions are as follows
\begin{eqnarray}
\label{53}
\mathbf{S}&=&
\begin{cases}
S_{x} =\sum\limits_{b=1,2} \sqrt{b}\left(|b\rangle\langle b-1|+|b-1\rangle\langle b|\right), \\
    
    S_y =\sum\limits_{b=1,2}\sqrt{b}\left(\rm{i}|b\rangle\langle b-1|-\rm{i}|b-1\rangle\langle b|\right),  \\     
    S_z =\sum\limits_{b=1,2,3}(3-2b)|b-1\rangle\langle b-1|.\\
\end{cases}
\end{eqnarray}

\begin{figure}[tbp]
  \centering
  \includegraphics[width=1\linewidth]{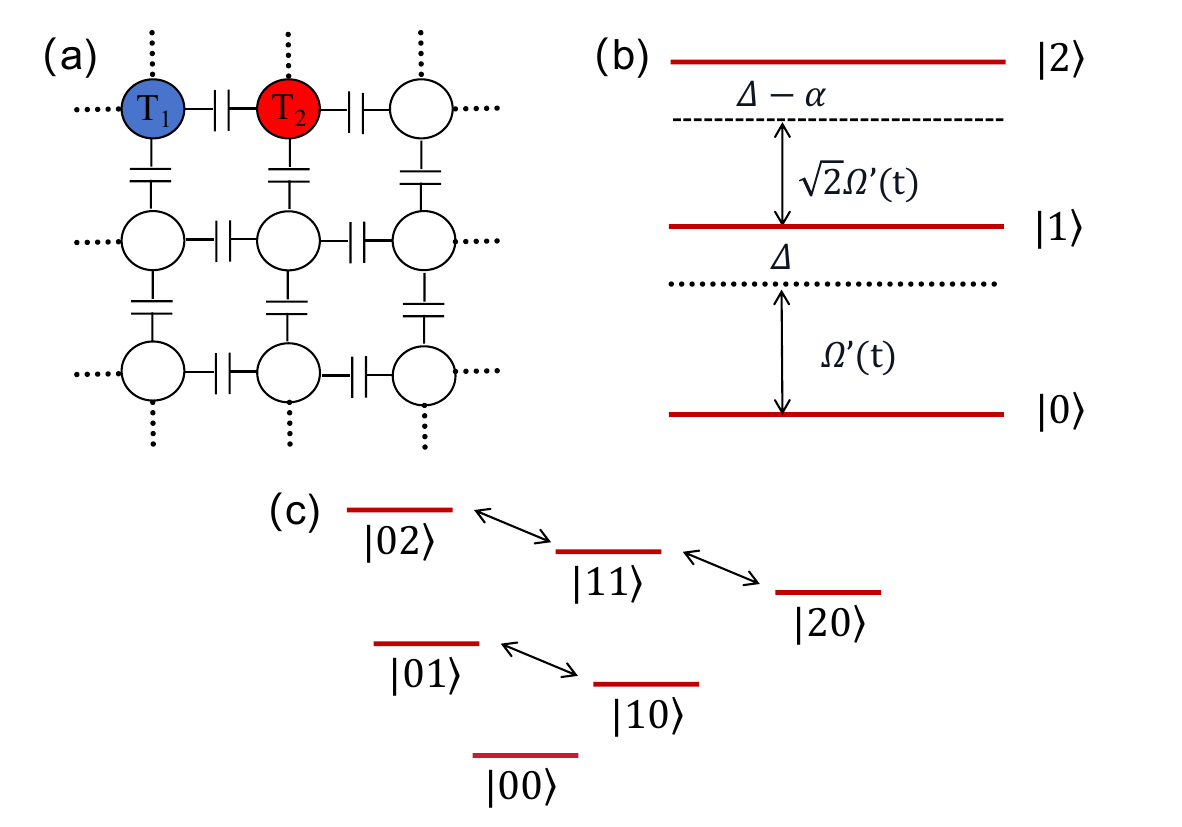}
\caption{Illustration of the  realization our scheme on superconducting circuits. (a) Schematic diagram of a 2D square  lattice with   capacitive coupled superconducting transmon qubits. (b) Illustration a  transmon qubit, with anharmonicity $\alpha$, driven by a microwave filed with   detuning $\Delta$. (c) Energy spectrum of two capacitively coupled transmons.}  
\label{Fig4}
\end{figure}

Through the above operation, $\mathcal{H}_D(t)$ in Eq. (\ref{67}) becomes
\begin{eqnarray}
\label{51}
\mathcal{H}_{D}(t)&=&\sum_{b=1}^{2}[\frac{1}{2}\Omega_D(t)\sqrt{b}|b-1\rangle\langle b|e^{-\rm{i}\phi(t)}+\rm{H.c.}] \notag\\
&&+\sum_{b=0}^{2}(b-\frac{1}{2})\Delta(t)|b\rangle\langle b|-\alpha|2\rangle\langle 2|,
\end{eqnarray}
where $\Omega_{D}(t)=\Omega'(t)-[{\rm i}\Omega'(t)+\Omega'(t)\phi(t)+\Delta\Omega'(t)]/(2\alpha)$ is the microwave pulse modified by DRAG technology. 

Next, we choose $H$ and $T$ gate as a reference to test the gate performance by selecting $c_{-}=|0\rangle\langle1|+\sqrt{2}|1\rangle\langle2|$, $c_{z}=|1\rangle\langle1|+2|2\rangle\langle2|$  in Eq. (\ref{EqMaster}), and replace $\mathcal{H}(t)$ with $\mathcal{H}_{D}(t)$ in Eq. (\ref{51}). In combination with current experimental techniques, we set $\alpha=2\pi\times 280$ MHz and $\kappa_1=\kappa_2=2\pi\times2$ kHz. Under the setting of these parameters, we change the maximum value of microwave amplitude $\Omega_{\rm{M}}$ to plot the change of the $H$ and $T$ gates fidelity with different values as Fig. 5(a), and show the dynamics of which in Fig. 5(b). When $\Omega_{\rm{M}}$ is equal to $2\pi\times51$ MHz, the $H$ gate fidelity has an optimal value of $99.97\%$. And the optimal fidelity of $T$ gate can exceed $99.98\%$ while $\Omega_{\rm{M}}$ is $2\pi\times38$ MHz.

In addition, we consider testing the state population and state fidelity with the same parameter settings. Given that the initial state $|\Psi_1(0)\rangle$ of the $H$ and $T$ gates are $|0\rangle$ and $(|0\rangle+|1\rangle)/\sqrt{2}$, respectively. The final state $|\Psi_1(\tau)\rangle$ becomes $(|0\rangle+|1\rangle)/\sqrt{2}$ and $[|0\rangle+\rm{exp}({\rm i}\pi/4)|1\rangle]/\sqrt{2}$ in an ideal case, and $F_s=\langle \Psi_1(\tau)|\rho_1|\Psi_1(\tau)\rangle$ is used to test the state fidelity. The resulting state population and fidelity dynamics are shown in Figs. 5(c) and 5(d), we can observe that the leakage to $|2\rangle$ is effectively suppressed, and both of the state fidelity can reach $99.97\%$. 

\begin{figure}[tbp]
  \centering
  \includegraphics[width=1\linewidth]{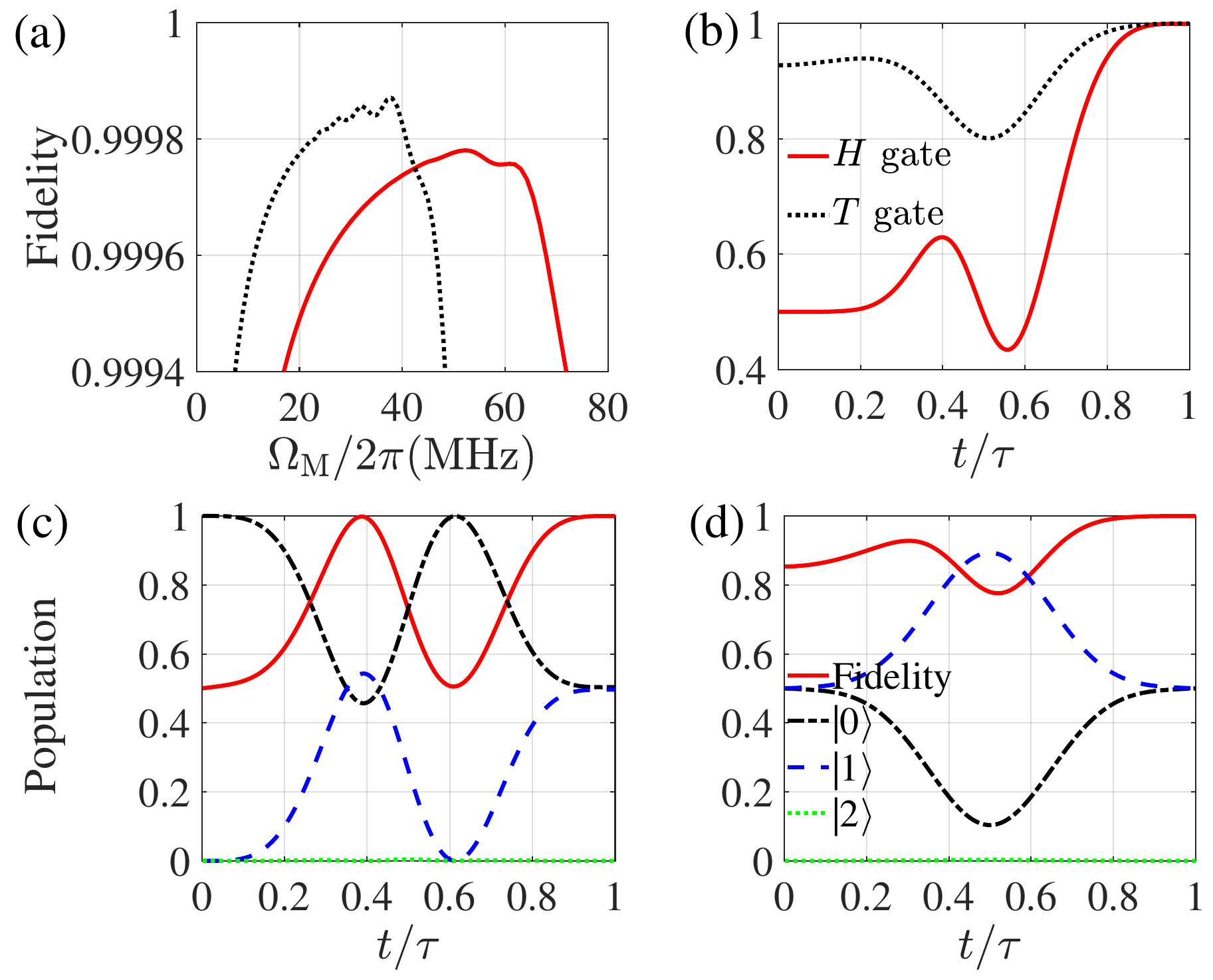}
\caption{(a) The change of gate fidelity with maximum value of microwave amplitude for $H$ and $T$ gates. (b) Dynamics of the $H$ and $T$ gate fidelity under the optimal $\Omega_{\rm M}$. State population and fidelity dynamics of (c) $H$ and (d) $T$ gates by setting the initial state $|0\rangle$ and $(|0\rangle+|1\rangle)/\sqrt{2}$. }
\label{Fig5}
\end{figure}

\subsection{ Nontrivial two-qubit geometric gate}

Now, two transmon qubits $T_1$ and $T_2$ are selected to realize the construction of nontrivial two-qubit geometric gate in the superconducting system. The coupled Hamiltonian between these two qubits can be written as
\begin{equation}
\label{54}
\begin{split}
\mathcal{H}_{12}=&\sum_{i=1}^{2}\sum_{j=1}^{\infty}[i\omega_i-\frac{j(j-1)}{2}\alpha_i]|j\rangle_i\langle j| \\
+&g_{12}[(\sum_{p=0}^{\infty}\sqrt{p+1}|p\rangle_1\langle p+1|\\
\otimes&\sum_{q=0}^{\infty}\sqrt{q+1}|q\rangle_2\langle q+1|)+\rm{H.c.}],
\end{split}
\end{equation}
where $g_{12}$ is the coupling strength between the two qubits, $\omega_i$ and $\alpha_i$ are the qubit frequency and anharmonicity, respectively. Since the coupling strength between $T_1$ and $T_2$ is fixed, we use the tunable coupling technology \cite{ Cai, Chu} to make it adjustable. We control the frequency of $T_1$ by adding ac driving to it, so that the frequency of $T_1$ becomes $\omega_1(t)=\omega_1+\beta\nu\sin[\nu t+\eta(t)]$. Transforming the above Eq. (\ref{54}) accordingly, and then ignoring the high frequency oscillation terms in the interaction picture, we can get
\begin{equation}
\label{55}
\begin{split}
\mathcal{H}'_{12}(t)=&g_{12}\{[|01\rangle\langle10|e^{{{\rm i}}\Delta_{12}t}+\sqrt{2}|02\rangle\langle11|e^{{\rm i}(\Delta_{12}-\alpha_2)t} \\
 +&\sqrt{2}|11\rangle\langle20|e^{{\rm i}(\Delta_{12}+\alpha_1)t}]e^{-{\rm i}\beta\sin[\nu t+\eta(t)]}\}+{\rm{H.c.}},
 \end{split}
\end{equation}
where $\Delta_{12}=\omega_2-\omega_1$ indicates the frequency difference between $T_1$ and $T_2$. Fig. 4(c) shows the corresponding energy spectrum of the two capacitively coupled transmons $T_1$ and $T_2$. After setting parameter $\Delta'=\nu-\Delta_{12}+\alpha_2$, a nonresonant transition occurs in the subspace $\{|02\rangle, |11\rangle\}$. Besides, use  $U_{\Delta'}={\rm exp}[-{\rm i}\Delta'(|02\rangle\langle02|-|11\rangle\langle11|)t]$ to make a transformation. Meanwhile, with  the help of the Jacobi–Anger equation ${\rm exp}({\rm i}\beta\cos\theta)=\sum_{h=\infty}^{\infty}{\rm i}^{h}J_h(\beta){\rm exp}({\rm i}h\theta)$, neglecting the high-order oscillating terms, we will obtain an two-level effective Hamiltonian 
\begin{equation}
\label{56}
\begin{split}
\mathcal{H}_e(t)=&\frac{1}{2}\left(
  \begin{array}{cc}
    -\Delta'& \Omega_{12}(t)\\
   \Omega_{12}^*(t)& \Delta'\\
\end{array}\right),\\
\end{split}
\end{equation}
where $\Omega_{12}(t)=2\sqrt{2}g_{12}J_1(\beta)e^{-i\eta(t)}$, and the intensity of microwave is determined by $\beta$. Obviously, the Eq. (\ref{56}) has the same structure as the Eq. (\ref{4}), so we can construct two-qubit control phase gates in the subspace $\{|00\rangle, |01\rangle, |10\rangle,|11\rangle\}$, and the corresponding evolution operator is
\begin{equation}
\label{57}
U_{12}=|00\rangle\langle00|+|01\rangle\langle01|+|10\rangle\langle10|+e^{{\rm i}\gamma'_g}|11\rangle\langle11|. 
\end{equation}

\begin{figure}[tbp]
  \centering
  \includegraphics[width=0.8\linewidth]{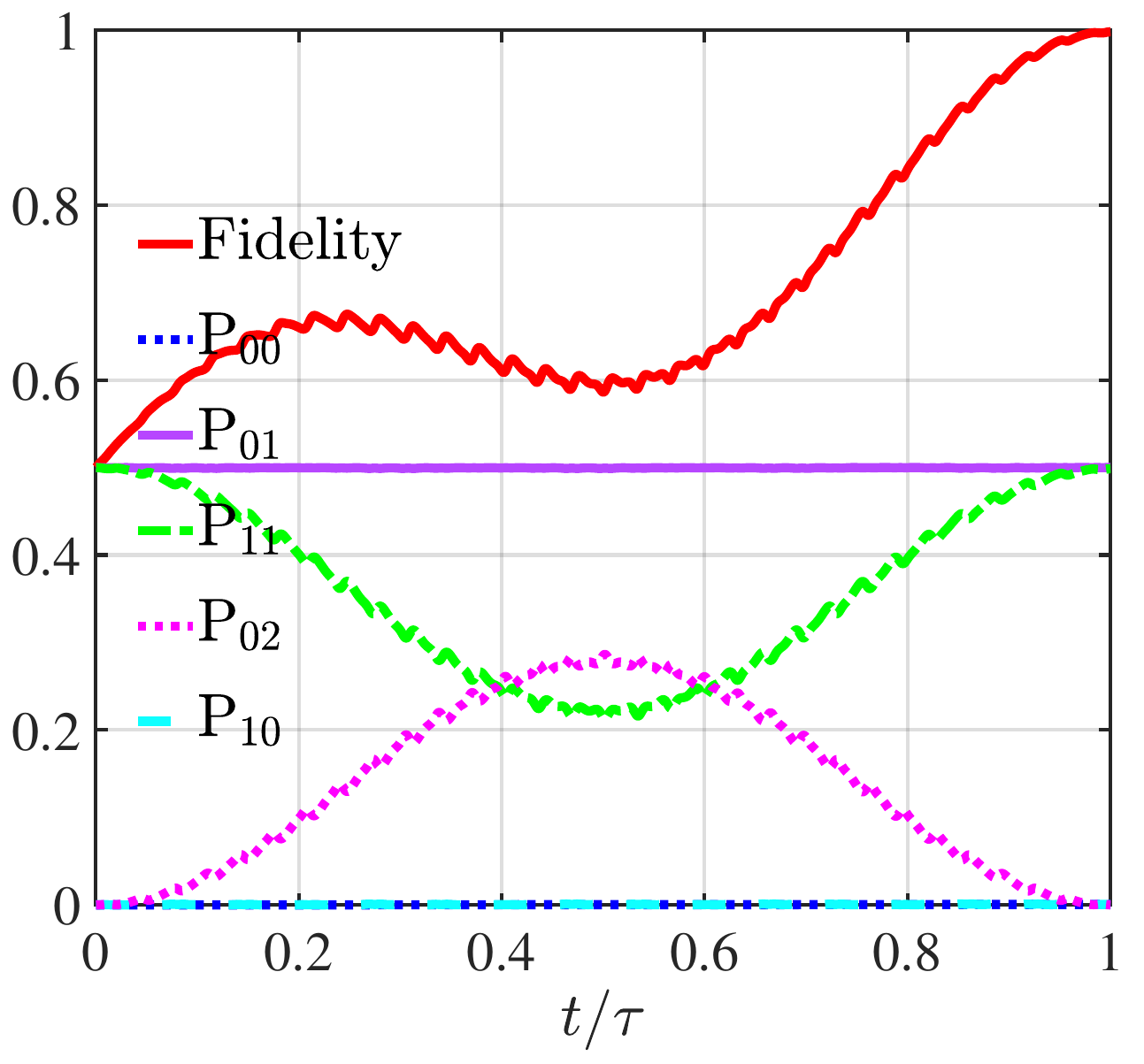}
\caption{  State population and fidelity dynamics for the two-qubit control phase gate, with $\gamma'_g= \pi/2$, by setting the initial state $\left(|01\rangle+|11\rangle\right)/\sqrt{2}$. See the maintext for detail parameters settings.}
\label{Fig6}
\end{figure}

Take $\gamma'_g= \pi/2$ as an example by setting parameters $\Gamma=\xi=0$ and $\varphi_{-}=3\pi$ in Eq. (\ref{12}). Considering the influence of environment-induced decoherence, we apply the master equation
\begin{equation}
\label{EqMaster'}
\begin{split}
\dot\rho_2=&-{\rm i}[\mathcal{H}'_{12}(t), \rho_2]+\kappa_{1}\chi(c_{-})+\kappa_{2}\chi(c_{z})\\
+&\kappa'_1\chi(c'_{-})+\kappa'_2\chi(c'_{z})
\end{split}
\end{equation}
to test. Here, $c'_{-}=|0\rangle_2\langle1|+\sqrt{2}|1\rangle_2\langle2|$ and $c'_{z}=|1\rangle_2\langle1|+2|2\rangle_2\langle2|$.
We use $F_2=\int_0^{2\pi}\int_0^{2\pi}\langle\Psi_2(\tau)|\rho_2|\Psi_2(\tau)\rangle d\Theta_1d\Theta_2/(2\pi)^2$ to test the gate fidelity, which is numerically integrated of 10001 initial states with $\Theta_1$ and $\Theta_2$  uniformly distributed between $0$ and $2\pi$. Set the parameters $g_{12}=2\pi\times5$ MHz and $\kappa_1=\kappa_2=\kappa'_1=\kappa'_2=2\pi\times2$ kHz, and then adjust the value of $\beta$ to reduce the damage of the gate fidelity caused by high-order oscillation. Combined with current experimental techniques, we finally selected $\beta=1.7$, $\Delta_{12}=2\pi\times600$ MHz, $\alpha_1=2\pi\times300$ MHz and $\alpha_2=2\pi\times280$ MHz. Under the setting of these parameters, $\Delta'\approx-2\pi\times6.9$ MHz. Then, we can get $\nu=\Delta'+\Delta_{12}-\alpha_2=2\pi\times313.1 
$ MHz. Given the initial state $|\Psi_2(0)\rangle=\left(|01\rangle+|11\rangle\right)/\sqrt{2}$, the ideal final state is obtained by using $|\Psi_2(\tau)\rangle=U_{12}|\Psi_2(0)\rangle$, and then the corresponding state population and state fidelity can be plotted. Fig. 6 shows that it is not much leakage and the state fidelity can reach $99.80\%$. Changing the initial state to $|\Psi_2(0)\rangle=\left(\cos\Theta_1|0\rangle_1+\sin\Theta_1|1\rangle_1\right)\otimes\left(\cos\Theta_2|0\rangle_2+\sin\Theta_2|1\rangle_2\right)$, we can get  the control phase gate fidelity is higher than $99.87\%$ under the same parameter setting.

\section{Discussion and conclusion}
In summary, our scheme is based on the design of multi-degree-of-freedom auxiliary vector to reversely solve the Hamiltonian and construct the NPGQC scheme. Since the scheme only accumulates geometric phases after the evolution of any initial state, the quantum gates are robust against some control errors. Our scheme highlights the advantages of geometric phase and improves the gate fidelity. It is worth noting that our scheme also implements rapid evolution through a noncyclic smooth evolution path which removes the segmented variation of the Hamiltonian during the process.
 
On the basis of that, we also realize the construction of universal single-qubit and nontrivial two-qubit control phase gates in superconducting physical system, whose gate fidelity can be respectively higher than $99.97\%$ and $99.87\%$. The results show that our scheme can improve the fidelity and robustness of the universal quantum gate set in practical applications of quantum computation. Meanwhile, our scheme can be readily extended to other systems, e.g., trapped ions, NV
centers in diamond and Rydberg atoms. Therefore,  our scheme representing  a promising step forwards  for practical quantum  computation.

\begin{acknowledgements}
This work was supported by the Key-Area Research and Development Program of GuangDong Province (Grant No. 2018B030326001),  the National Natural Science Foundation of China (Grant No. 12275090) and Guangdong Provincial Key Laboratory (Grant No. 2020B1212060066).
\end{acknowledgements}

\end{document}